# Identifying the magnetospheric driver of STEVE


**Xiangning Chu[1], David Malaspina[1], Bea Gallardo-Lacourt[2], Jun Liang[2], Laila Andersson[1], Qianli Ma[3], Anton Artemyev[4], Jiang Liu[3], Bob Ergun[1], Scott Thaller[1], Hassanali Akbari[1], Hong Zhao[1], Brian Larsen[5], Geoffrey Reeves[5], John Wygant[6], Aaron Breneman[6], Sheng Tian[6], Martin Connors[7], Eric Donovan[2], William Archer[8], Elizabeth A. MacDonald[9]**

[1] Laboratory for Atmospheric and Space Physics, University of Colorado Boulder, Boulder, Colorado, USA

[2] Department of Physics and Astronomy, University of Calgary, Calgary, Alberta, Canada

[3] Department of Atmospheric and Oceanic Sciences, University of California, Los Angeles, California, USA

[4] Institute of Geophysics and Planetary Physics, University of California, Los Angeles, CA, USA

[5] Los Alamos National Laboratory, Los Alamos, NM, United States

[6] Department of Physics, University of Minnesota, Twin Cities, Minneapolis, MN, USA

[7] Athabasca University, Athabasca, Canada

[8] Department of Physics and Engineering Physics, University of Saskatchewan, Saskatoon, Saskatchewan, Canada

[9] NASA Goddard Space Flight Center, Greenbelt, MD, USA

Corresponding author: Xiangning Chu (chuxiangning@gmail.com)




**Key Points:**

- First identification of the magnetospheric driver of STEVE using conjunction magnetospheric observations

- STEVE's driver region features strong quasi-static electric field, broadband waves, parallel electron acceleration, and perpendicular ion drift

- Ionospheric electron heating or Alfven wave accelerated electrons could power continuous STEVE emission

**Time line:**

November 20, 2018      Research PowerPoint sent to coauthors

December 12, 2018      Research presented at AGU Fall meeting 2018

January 14, 2019      Manuscript sent to coauthors

March 11, 2019      Manuscript submitted to GRL



**Abstract**

For the first time, we identify the magnetospheric driver of STEVE, east-west-aligned narrow emissions in the subauroral region. In the ionosphere, STEVE is associated with subauroral ion drift (SAID) features of high electron temperature peak, density gradient, and strong westward ion flow. In this study, we present STEVE's magnetospheric driver region at a sharp plasmapause containing: strong tailward quasi-static electric field, kinetic Alfven waves, parallel electron acceleration, perpendicular ion drift. The observed continuous emissions of STEVE are possibly caused by ionospheric electron heating due to heat conduction and/or auroral acceleration process powered by Alfven waves, both driven by the observed equatorial magnetospheric processes. The observed green emissions are likely optical manifestations of electron precipitations associated with wave structures traveling along the plasmapause. The observed SAR arc at lower latitudes likely corresponds to the formation of low-energy plasma inside the plasmapause by Coulomb collisions between ring current ions and plasmaspheric plasma.

**1 Introduction**

The aurora is a spectacular natural phenomenon that occurs in the polar regions. It occurs statistically within the auroral oval, an oval-shaped belt around the magnetic pole (usually between 65-80º in magnetic latitude) [*Feldstein*, 1963]. Auroral features are caused by the dynamics of energetic particles in the Earth's magnetosphere as they precipitate into the upper atmosphere. There are many types of ionospheric photon emissions, which are generally generated by the impact excitation by 1) precipitating electrons followed by the de-excitation of the neutrals releasing the photon such as aurora or 2) local ionospheric process such as stable red auroral (SAR) arc. Their electron source and energy are vastly different. There are two types of



classic aurora, the diffuse and the discrete aurora. Diffuse aurora is powered by energetic electron precipitation as electrons are scattered into the loss cone via wave-particle interactions [*Ni et al.*, 2008; *Thorne et al.*, 2010]. Discrete aurora is caused by precipitating electrons between 1-10 keV accelerated by quasi-static field-aligned potential drop located several thousand km or even a few $R_E$ above the Earth [*Hull et al.*, 2003; *Mozer et al.*, 1977]. The electrons could also be accelerated to a few hundred eV by time-dependent Alfven waves with strong Poynting flux [*Chaston et al.*, 2003; *Ergun et al.*, 2001; *Keiling et al.*, 2003; *Lysak and Song*, 2003; *Wygant et al.*, 2002]. In addition to precipitation, local ionospheric processes cause emission classified as airglow (often referred as dayglow or nightglow [*Solomon*, 2017] and reference therein), or SAR arc by heat flux from the magnetosphere [*Barbier*, 1958; *Hoch and Lemaire*, 1975; *M. Rees and Roble*, 1975; *Shiokawa et al.*, 2013].

A distinctive emission feature, STEVE (Strong Thermal Emission Velocity Enhancement), was reported for the first time by the scientific community [*MacDonald et al.*, 2018] with the help of citizen scientists (amateur auroral photographers) via Aurorasaurus [*MacDonald et al.*, 2015]. It occurs in the subauroral region equatorward of the auroral oval and it is thin in latitude and wide in longitude. In addition, STEVE has green picket-fence auroral features which propagate westward along the purple arc. In *MacDonald et al.* [2018], a STEVE event was studied using conjunction observations from the Redline Emission Geospace Observatory (REGO) all-sky imager (ASI) at Lucky Lake (LUCK) and a Low Earth Orbit (LEO) satellite Swarm-B. In the ionosphere, STEVE was accompanied by increased electron temperature, depleted plasma density, strong westward ion velocity and weak magnetic perturbations indicative of field-aligned currents (FACs), which are consistent with the ionospheric features of an SAID. A recent study, using observations from the LEO satellite



POES-17, showed that STEVE might not be associated with particle precipitation, suggesting that STEVE might be generated by local ionospheric process [*Gallardo-Lacourt et al.*, 2018]. However, the lowest energy channel of the particle detectors onboard POES-17 measured the total flux integrated between 50-2000 eV, therefore it did not rule out the possibility that STEVE might be associated with lower-energy ions or electrons. As POES-17 lacks in-situ thermal plasma measurement capability, the ionospheric process producing STEVE optical emission remains unclear. After the second revision of our paper was sent in for review, an independent study of the same topic appeared [*Nishimura et al.*, 2019]. The main difference between is that their magnetospheric observations are conjunction to the dark ionosphere about 1-hour magnetic local time away from STEVE; our magnetospheric satellites' footprint crossed STEVE directly.

In addition to unknown ionospheric processes, the magnetospheric driver powering STEVE is unknown. In this study, we identify the magnetospheric driver of STEVE using conjunction observations from ground-based ASIs, LEO satellite, and magnetospheric spacecraft. Our study reveals that the STEVE was associated with a large quasi-static electric field at a sharp plasmapause boundary. Strong kinetic Alfven waves, parallel electron acceleration, and perpendicular ion drift were observed at the plasmapause, providing several explanations for the underlying causes of STEVE.

## 2 Observations

We report a STEVE event occurred on 17 July 2018 in conjunction with ground-based ASIs, Swarm-B satellite, and two Van Allen Probes. The STEVE event occurred during a 2-hour period of continuous activity, which is consistent with a previous study [*Gallardo-Lacourt et al.*, 2018]. The Sym-H index started to decrease around 0430 UT and reached -30 nT around 0730



UT. The THEMIS AL index started to drop around 05:20 UT and reached a minimum of -600 nT.

## 2.1 Ground-based optical observations

STEVE was captured by two white-light THEMIS ASIs at Athabasca (ATHA) and Pinawa (PINA) between 06:30 UT to 07:20 UT on 17 July 2018. Figure 1 shows the long and narrow band of the continuous STEVE emission at four selected times (see movie S1 for the entire image sequence). The continuous emission of STEVE was captured by a REGO ASI at LUCK (the same four times in Figure S1). In addition, both the continuous narrow 'purple' emission and green picket-fence structures were captured by two citizen scientists, from both northern and southern sides of STEVE (photographs by Colin Chatfield and Neil Zeller are available in Figure S2). The purple and green emissions overlapped near the zenith (Figure S2b), suggesting that two emissions were aligned nearly along the magnetic field lines. The continuous emission was located higher in altitude than the green emission when viewed away from the zenith (Figure S2a). STEVE was ~ 0.2° wide in magnetic latitude (MLAT) and at least 2 hours wide in magnetic local time (MLT) (limited by the field of view of the ASIs). STEVE first occurred at about 61° MLAT, gradually moved equatorward to ~59.5° MLAT, and faded away afterward. In addition to STEVE, the REGO ASI at LUCK also observed an east-west aligned SAR-like arc, which was only visible in the redline emission. The observations show that the SAR-like arc was different from STEVE during this event (Figure 1 and S1), which is clear when watching two movies S1 and S2 side-by-side. 1) The SAR-like arc was only visible in the redline emission but not in the white-light ASIs; 2) it occurred earlier than STEVE; 3) it was thicker and was located at a lower latitude (see Figure S1c and Figure 3b). Thus, the observations from white-light THEMIS ASIs and photographs from citizen scientists are essential



supplementary to identify STEVE from the SAR-like arc. Their differences in timing, duration, morphology, and types of emissions also suggest that STEVE and the SAR-like arc have different generation mechanisms. With the overlaps in the field-of-views between the ASIs, the STEVE peak emission altitude was determined using traditional triangulation method to be ~ 280 km. This STEVE altitude corresponds to the altitude of maximum ionization rate of low energy electrons (<100 eV) [*Manfred Rees*, 1964]. Uncertainties of the altitude estimate could be contributed from many factors and require further investigation.

The proton aurora, equatorward of the discrete aurora, was observed by Forty-Eight Sixty-One (FESO) meridian scanning photometers (MSPs) that record proton aurora emission at 486.1 nm [*Unick et al.*, 2017] at LUCK and Athabasca University. The proton aurora was located a few degrees poleward of STEVE during this event (see Figure S3). Therefore, STEVE did not correspond to the classic proton aurora or discrete electron auroral arcs (further poleward than proton aurora) [*Davis*, 1978; *Yahnin et al.*, 1997], consistent with *MacDonald et al.* [2018].

**2.2 Ionospheric observations: Swarm**

Ionospheric variations associated with STEVE were observed by Swarm-B in LEO. The Swarm constellation consists of three identical satellites [*Friis-Christensen et al.*, 2006]. Swarm-B has an altitude of ~530 km, while the other two satellites fly almost side-by-side at ~480 km. Each Swarm satellite carries high-precision and high-resolution instruments [*Knudsen et al.*, 2017; *Ritter et al.*, 2013] to measure magnetic field, ion flow velocity, ion and electron temperatures, and plasma density.

Swarm-B crossed east of STEVE at ~06:35:45 UT (red line in Figure 1). Observed ionospheric features (Figure 2) were consistent with an SAID [*MacDonald et al.*, 2018], which was located at the poleward edge of the subauroral region. The field-aligned currents (Figure 2a),



calculated from observed magnetic perturbations, were ~1.0 $\mu A/m^2$ narrow upward (negative) at

higher latitudes and ~1.0 $\mu A/m^2$ thicker downward (positive) at lower latitudes. The FACs were

associated with the eastward current on the plasmapause (see section 2.3). The electron density

(Figure 2b) showed a sharp gradient from $1.3 \times 10^4$ $cm^{-3}$ at lower latitudes to $4.0 \times 10^4$ $cm^{-3}$ at

higher latitudes. The ion velocity (Figure 2d) showed a peak at the STEVE latitude, reaching a

westward speed of 3 km/s corresponding to a strong poleward electric field. The electron

temperature (Figure 2c) increased from 2500 K outside to 7600 K at its center, which was

extremely high. Electron heating to a few thousand Kelvins is known to produce red auroral

emission at 630.0 nm by exciting the lowest electronic state in atomic oxygen O(1D), which is

responsible for stable red auroral (SAR) arcs [*Förster et al.*, 1999; *Foster et al.*, 1994; *Sazykin et

al.*, 2002]. The above features are consistent with the signatures of SAID in the ionosphere

[*Anderson et al.*, 1993; *Galperin*, 2002; *He et al.*, 2014; *Mishin*, 2013; *Puhl-Quinn et al.*, 2007;

*Spiro et al.*, 1977].

### 2.3 Magnetospheric observations: Van Allen Probes

In this study, the magnetospheric driver of STEVE is identified using the observations

from the Van Allen Probes as shown in Figure 3. Van Allen Probes are two identical spacecraft

in nearly the same highly elliptical, low inclination orbits ($1.1 \times 5.8$ $R_E$, 10º) [*Mauk et al.*, 2012].

During this event, two Van Allen Probes traveled tailward, from the plasmasphere into the

plasma sheet at pre-midnight local time (see *L* shell in the time axis labels). Their magnetic

footprints crossed STEVE successively around 06:53 UT and 06:47 UT at two separated MLTs

(Figure 1). As their footprints traversed STEVE (indicated by the luminosity peaks at their

footprints, Van Allen Probe B shown), they crossed the plasmapause (PP) where the plasma

density (Figure 3c) decreased from ~1000 $cm^{-3}$ to 10 $cm^{-3}$ within 500 km (<0.1 $R_E$). The plasma



density is inferred from the upper hybrid resonance frequency observed by the EMFISIS

instrument [*Kurth et al.*, 2015]. In addition, their footprints crossed the SAR arc earlier at lower

latitudes (Figure 3b).

STEVE is associated with an SAID at the plasmapause. The background magnetic field

was mainly northward in GSM coordinates with an amplitude of ~270 nT. The magnetic

perturbations (Figure 3d, subtracted by the IGRF model and detrended with a 20-min window)

showed a bipolar pattern in the northward component (Bz) with a peak-to-peak variation of ~20

nT at the plasmapause suggesting the existence of an eastward current. A strong tailward quasi-

static electric field of ~-20 mV/m (Figure 3e) was observed at the plasmapause by Van Allen

Probe-B (the DC-coupled electric field and other measurements on Van Allen Probe-A were

similar, but they were not reliable during this event due to radiation damage accumulated by the

preamplifiers and thus not shown). This strong electric field corresponded to a strong westward

plasma drift velocity $V_{EXB}$ of ~ 80 km/s. The strong and narrow electric field is a typical feature

of an SAID in the equatorial magnetosphere ([*Mishin*, 2013] and references therein). The

observed SAID coincided with the inner edge of the ion plasma sheet, which is similar to

subauroral polarization streams (SAPS) [*Foster and Vo*, 2002; *Southwood and Wolf*, 1978; *Yeh

et al.*, 1991]. This eastward current and the ionospheric FACs were likely caused by the

misalignment of gradient in the ring current pressure (see last discussion paragraph) and

magnetic flux volume [*Vasyliunas*, 1970; *Wolf*, 1983; 1995]. The FACs flow into regions of low

ionospheric conductivity (low ionospheric electron density observed by Swarm-B), which lead to

the observed large electric field in order to maintain current continuity [*Foster and Burke*, 2002].

In this event, the ion and electron energies near the plasmapause reached a few hundred

volts as observed by the HOPE instrument [*Funsten et al.*, 2013] simultaneously with the strong



quasi-static electric field. The spacecraft floating potential of Van Allen Probe-B was only a few volts, suggesting that the observed plasma energy was real. The proton energy increased from a few eV to a few hundred eV at the peak of the electric field, and then dropped back. These protons were mostly perpendicular to the magnetic field, since the perpendicular proton flux was much higher than the parallel flux. The purple line in Figure 3g, which is the energy of protons moving at the speed of $V_{EXB}$ drift ($1/2m_p V_{EXB}^2$), agrees well with the observed proton flux peak energy. This fact suggests that the protons were drifting at the electric field drift velocity (not stochastic heating). Since the proton drift velocity at the plasmapause was perpendicular, it is unlikely that they precipitated into the ionosphere.

The electrons at the plasmapause were accelerated to a few hundred eV. The electron acceleration, on the other hand, was stochastic and mostly in the parallel direction since the normalized parallel electron flux increased significantly (Figure 3i, the ratio of the parallel electron flux to the omni-directional flux in Figure 3h). This parallel electron acceleration was associated with strong kinetic Alfven waves (KAW) at the plasmapause (Figure 3j-l). The electric and magnetic field wave power spectra from EMFISIS were rotated into local field-aligned coordinates. $Z_{FAC}$ is along the magnetic field B, $Y_{FAC}$ is perpendicular to both the spin axis and B (roughly westward), and $X_{FAC}$ completes the orthogonal set (roughly sunward). The power spectra in $E_{YFAC}$ and $B_{XFAC}$ (Figure 3j and 3k) show strong broadband waves, and the ratio of $E_{YFAC}/B_{XFAC}$ (Figure 3l) indicates that the Alfven waves were Doppler-shifted by the drift velocity $V_{EXB}$, which is a typical signature of KAW [*Chaston et al.*, 2014]. The black line in Figure 3i represents the maximum possible electron energy due to acceleration by the parallel electric field of the KAW [*Artemyev et al.*, 2015; *Kletzing*, 1994]. The fact that the parallel accelerated electron flux stayed below the black line suggests that the parallel electrons were



accelerated by the local kinetic Alfven waves. Note that the electron acceleration was underestimated because the integration did not include waves powers below the low-frequency cutoff of EMFISIS (4 Hz), which explains why the black line was lower coinciding with the KAW. These parallel low energy electrons (< a few hundred eV), if their pitch angles were within the local loss cone (7°), may precipitate into the ionosphere and contribute to the generation of STEVE. However, the pitch angle resolution of the HOPE instrument is too large (18°) to resolve the loss cone and directly calculate the energy flux of the electron precipitation. We have estimated an upper limit to the energy flux of the electron precipitation of 0.068 erg/cm$^2$/s when mapped to the ionosphere using the method developed by [*Mozer*, 1970]. Yet it is too weak to generate visible aurora which requires ~1 erg/cm$^2$/s [*Kivelson and Russell*, 1995]. There were substantially enhanced low-energy parallel electrons (~1-10 eV), however, which could transport heat flow into the topside ionosphere and result in electron heating during this event [*Evans and Mantas*, 1968; *Liang et al.*, 2017; *Moffett et al.*, 1998; *M. Rees and Roble*, 1975; *Slater et al.*, 1987]. In addition, the electromagnetic energy of the KAW propagated along magnetic field lines and could accelerate the precipitating electrons through auroral acceleration process close to the Earth [*Chaston et al.*, 2003; *Keiling et al.*, 2003; *Lysak and Song*, 2003; *Tian*, 2017; *Wygant et al.*, 2002], which could contribute to the generation of STEVE emissions (see discussion).

Inside the plasmapause, the protons (Figures 3f and 3g) were heated stochastically to tens of eV in both parallel and perpendicular directions. The electrons flux increased to a few hundred eV, suggesting that the entire region had substantially enhanced 1-10 eV electrons. In addition, the protons with tens of keV plasma sheet energy were observed inside the plasmapause, and their energies decreased as they drifted across the tailward electric field region at the



plasmapause, which was consistent with a ring current pressure gradient. These plasma

populations inside the plasmapause coincided with the observed SAR arc, which is consistent

with previous studies that the energy source of SAR arcs was related to the energy transfer from

ring current ions to plasmaspheric cold electrons and ions as a result of energy degradation

through Coulomb collisions [*Fok et al.*, 1993; *Kozyra et al.*, 1987].

## 3 Discussion

In this study, we identified the magnetospheric driver of STEVE using conjunction

observations from ground-based ASIs, Swarm-B, and the Van Allen Probes. An east-west

aligned narrow and long STEVE was observed in the subauroral region. Swarm-B observations

showed that STEVE was associated with a narrow region of strong westward ion drift

corresponding to strong poleward electric field, electron temperature peak, and a bipolar pattern

of FACs, which were ionospheric signatures of an SAID. Van Allen Probes observed tailward

electric fields near the plasmapause with substantially enhanced ions and electrons, and this

region seems to encompass both SAR arcs and STEVE. The magnetospheric driver of STEVE

was located at the sharp plasmapause, which was at the poleward/outer edge of this region. The

observed narrow, strong and quasi-static electric field corresponded to westward plasma drift,

which are magnetospheric features of an SAID. The electric field coincided with strong KAW,

parallel acceleration of electrons, perpendicular acceleration of ions up to a few hundred eV, and

energy discontinuity of penetrating plasma sheet plasma. The protons were drifting

perpendicularly, while the electron acceleration by the KAWs was stochastic and parallel. In

addition to STEVE, SAR arcs were observed at slightly lower latitudes and corresponded to the

regions inside the plasmapause, where ring current ions and enhanced low-energy plasma were

observed. This is consistent with previous studies that the energy source of SAR arcs is related to



the energy transfer from ring current ions to plasmaspheric cold electrons and ions as a result of energy degradation through Coulomb collisions [*Fok et al.*, 1993; *Kozyra et al.*, 1987].

The conjunction observations provide solid evidence that the magnetospheric driver of STEVE is an SAID at a sharp plasmapause. In this event, the morphological characteristics of STEVE gave clues as to its magnetospheric driver. Its occurrence at the poleward/tailward edge of the active subauroral region was consistent with the fact that its driver, an SAID, occurred at the plasmapause earthward of the plasma sheet. It was narrow in latitude and extended several hours in MLT, consistent with the fact that the SAID electric field at the plasmapause was radially narrow and azimuthally broad enough to be seen at different MLTs by two Van Allen Probes and Swarm-B. STEVE lasted about one hour, which was consistent with the duration of an SAID ([*Mishin*, 2013] and references therein). Besides, STEVE was different from the SAR arc in location, time and wavelength during this event. Therefore, although it has been suggested that SAIDs might be related to SAR arcs [*Förster et al.*, 1999; *Foster et al.*, 1994], the unique characteristics of STEVE suggest that the SAIDs should be related to STEVE in this event.

Both the strong continuous narrow 'purple' emission and green picket-fence structures were captured by citizen scientists. STEVE's morphological features and different emission wavelengths suggest the presence of multiple underlying mechanisms discussed below.

The continuous STEVE emission could be contributed by the redline emission due to the ionospheric electron heating (similar to SAR arc). Previous studies suggest that redline emission could be many kR for Te > 4000 K [*Carlson et al.*, 2013; *Kozyra et al.*, 1990]. During the STEVE event, the ionospheric electron temperature reached a peak of ~7600 K. The ionospheric electron heating is usually caused by the heat conduction from the magnetosphere [*Liang et al.*, 2017; *Sazykin et al.*, 2002]. More specifically for this STEVE event, by the stochastic electron



acceleration coincided with the SAID. Such a high electron temperature could contribute to the redline emission of STEVE observed by REGO at LUCK, which is estimated below. The volume emission rate at 630.0 nm as a line-of-sight integral from [*Carlson et al.*, 2013] is:

$$I_{630} = \int_{250\ km}^{650\ km} \alpha(Te(h)) No(h) Ne(h)\, dh \tag{1},$$

where $\alpha(Te) = 0.15\sqrt{Te}\ \frac{(8537+Te)}{(34,191+Te)^3} \exp\left(-22{,}756/Te\right)$ is the O($^1$D) thermal electron excitation rate in cm$^3$/s, oxygen density No(h) and electron density Ne(h) are in cm$^{-3}$ and temperature Te in Kelvin. Ne is the parameter of least impact on the integrated intensity of thermal 630.0 nm emission, and Te clearly dominates [*Carlson et al.*, 2013]. Since the observations were insufficient to directly resolve the emission rate, we have compared our event with a SAR arc reported by *Foster et al.* [1994]. The red arc with 350 R brightness reported by *Foster et al.* [1994] corresponded to Te of 3500 K to 4000K, and Ne ~ 2.0 x10$^4$ cm$^{-3}$ at 450 km (slightly higher than 1.3x10$^4$ cm$^{-3}$ at 530 km for the current event). Since Te is the dominating parameter, Te ~ 7600 Kelvin could have produced red auroral emission of 7 to 17 kR which is visible to the human eye. Note that the emission may be somewhat overestimated because the ionospheric electron density is usually more depleted within a strong ionospheric flow. Nevertheless, the redline emission during this STEVE event is strong due to the extreme electron temperature.

The continuous STEVE emission could also be contributed by electron precipitation accelerated to a few hundred eV by the auroral acceleration process powered by the Alfven waves. Precipitating electrons into the atmosphere can be accelerated by two mechanisms: quasi-static potential drops and time-dependent wave activity. In this study, observations at appropriate altitudes and latitude were not available to analyze if any parallel potential drop developed. It



was suggested that STEVE is not dominated by precipitating electrons with energies of over a few keV [*Gallardo-Lacourt et al.*, 2018; *MacDonald et al.*, 2018]. Our measurements also showed that the estimated energy flux of the equatorial precipitating electrons was too weak to produce any visible emission. On the other hand, the observed Alfven waves were able to power the auroral acceleration process associated with field-aligned electron acceleration to a few hundred eV. This is consistent with the electron energy (<100 eV) corresponding to the maximum ionization at the STEVE altitude [*Manfred Rees*, 1964]. Therefore, although the energy flux of the parallel electrons at the equator was weak, they may be accelerated in the auroral acceleration region by strong Alfven waves and contributed to the continuous STEVE emission.

The westward traveling green picket-fence of STEVE were clearly related to strong waves traveling along the plasmapause. The lifetime of the green line auroral emission (< 1 sec) is shorter than the redline emission (~tens of sec) [*Gustavsson et al.*, 2001]. Therefore, the green emission is likely an instantaneous optical manifestation of electron precipitations of ~keV energy produced by strong Alfven waves; whereas the redline emission is the accumulation of these waves. Previous studies have suggested the existence of ballooning instability, shear-flow instability [*Kelley*, 1986; *Lakhina et al.*, 1990], interchange stability [*Sazykin et al.*, 2002], or Kelvin-Helmholtz instability [*Lui et al.*, 1982] at the plasmapause associated with density gradient and/or azimuthal flows. Instabilities are clearly at work at the sharp plasmapause where strong density and temperature gradients co-existed with strong azimuthal flow shear. Which instabilities are active is a challenging question and will require further investigation with more conjugate observations.



## 4 Conclusion

We identified the magnetospheric driver of STEVE using conjunction observations from ground-based ASIs, LEO Swarm-B satellite and two Van Allen Probes. STEVE was located at the poleward/tailward edge of an active subauroral region with strong poleward electric fields in the ionosphere with sharp electron density, downward FAC and tailward electric field near the plasmapause with substantially enhanced low energy electrons. This region seems to encompass both the SAR arc and STEVE. STEVE appears to be the optical manifestation of an SAID at the sharp plasmapause: a narrow quasi-static electric field maps along magnetic field lines, coinciding with significant plasma acceleration and strong KAW. The ionospheric electron heating due to heat conduction from the magnetosphere could have generated visible red aurora and contributed to the continuous STEVE emission. Parallel electron acceleration through auroral acceleration processes powered by Alfven waves could also contribute to the continuous emission. The green picket-fence emission of STEVE was likely related to electron precipitation caused by the waves traveling westward along the plasmapause. The SAR arc observed at lower latitudes likely corresponded to the formation of low-energy plasma population inside the plasmapause due to Coulomb collisions between ring current ions and plasmaspheric cold population. These ionospheric features and magnetospheric structures related to STEVE and SAR arc show strong implications to the inner magnetosphere-ionosphere coupling processes.

## Acknowledgments and Data

This work was supported by NASA awards NNX17AB81G and NAS5-01072. The authors thank Robert McPherron, Jacob Bortnik, and Jinxing Li for helpful discussions. The authors thank Colin Chatfield and Neil Zeller for their STEVE photographs and Johnathan Burchill for assistance in calibrating the Swarm ion drift measurement. We acknowledge



NSSDC Omniweb for geomagnetic activity indices (spdf.gsfc.nasa.gov). The authors thank the Van Allen Probes team, especially the EFW, EMFISIS, and ECT teams for their support (rbspgway.jhuapl.edu/data_instrumentationSOC). Swarm data is available from swarm-diss.eo.esa.int. THEMIS ASI data is available from themis.ssl.berkeley.edu. REGO ASI and FESO data is available from data.phys.ucalgary.ca. Data processing was done using SPEDAS [*Angelopoulos et al.*, 2019].



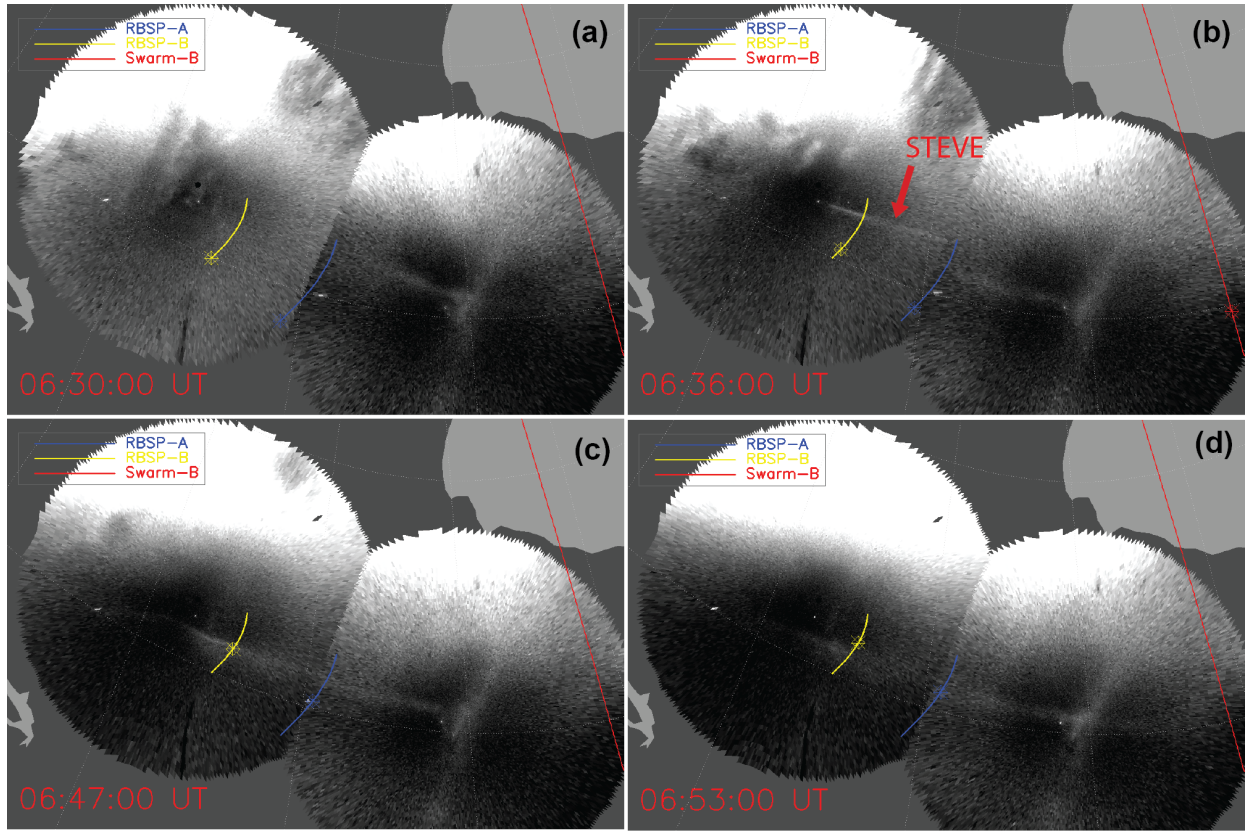

**Figure 1**. The STEVE event observed on 17 July 2018 by the THEMIS ASI at ATHA and PINA at (a) 06:30:00, (b) 06:36:00, (c) 06:47:00 and (d) 06:53:00 UT, with the footprints of Van Allen Probes and Swarm-B.



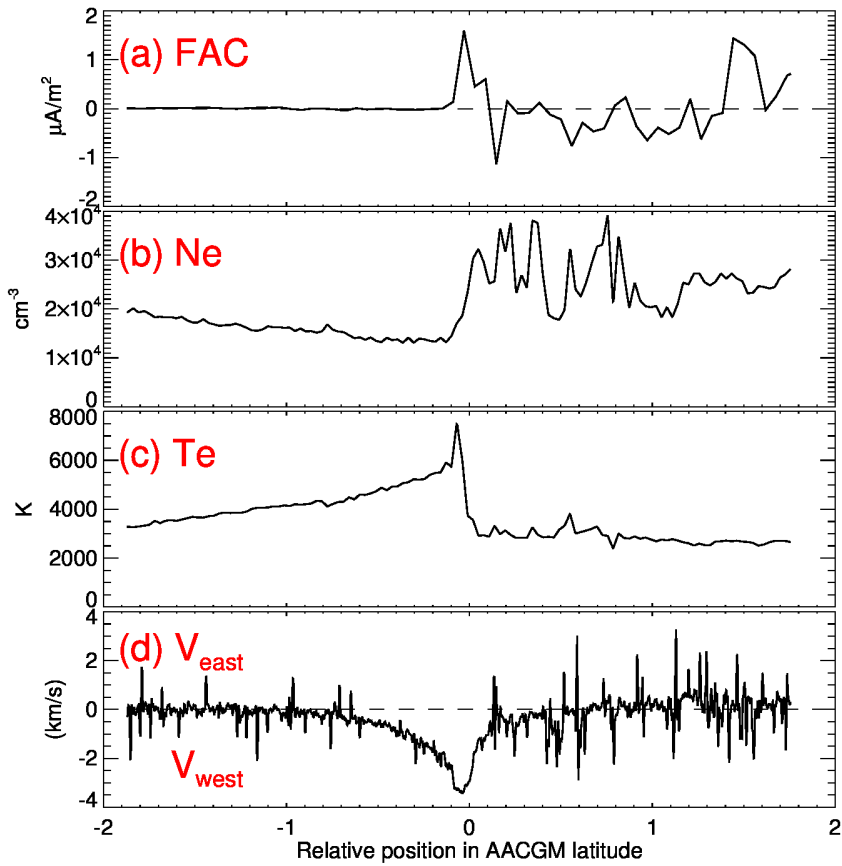

**Figure 2**. Ionospheric observations of STEVE versus relative latitude when Swarm-B crossed east of the STEVE (06:35:45 UT). The panels show (a) field-aligned currents, (b) electron density, (c) electron temperature, and (d) ion drift (positive eastward and negative westward).



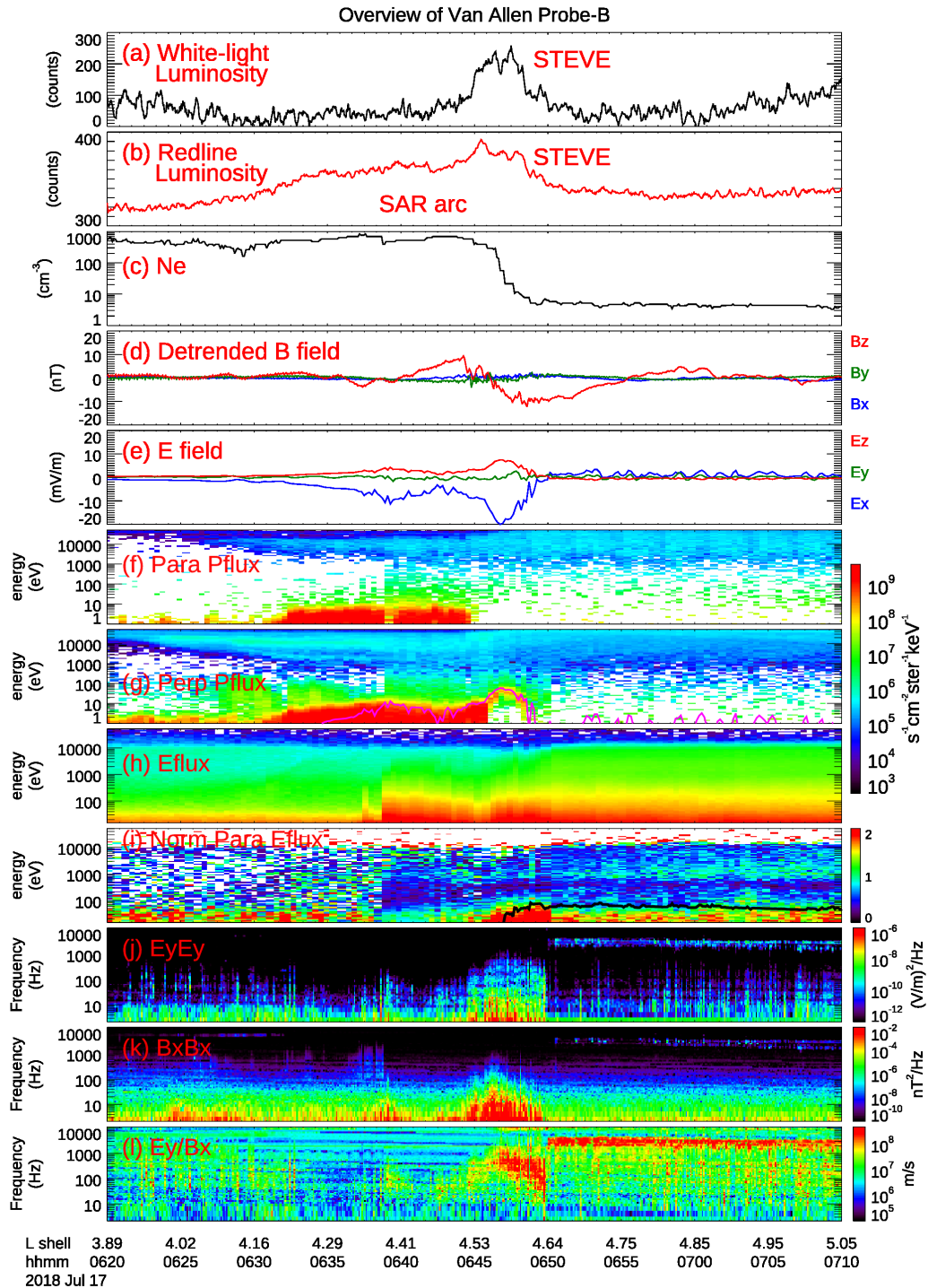

**Figure 3**. Overview of magnetospheric observations. (a) Auroral luminosities at the footprints

from white-light ASIs; (b) redline auroral luminosities at the footprints; (c) electron density; (d)



detrended magnetic fields in GSM coordinates; (e) electric field vectors in mGSE coordinates; (f) parallel differential proton flux at PA=18º; (g) perpendicular differential proton flux at PA=90º; the purple line represents the proton energy ($1/2m_pV_{EXB}^2$) at the speed of $V_{EXB}$ drift; (h) omnidirectional electron flux; (i) normalized parallel differential electron flux at PA=18º; the black line represent the electron energy accelerated by the KAW; (j) and (k) power spectra of the electric field ($E_{YFAC}$) and the magnetic field ($B_{XFAC}$) in FAC coordinates; (l) the ratio of $E_{YFAC}/B_{XFAC}$.



# References


Anderson, P. C., W. B. Hanson, R. A. Heelis, J. D. Craven, D. N. Baker, and L. A. Frank (1993), A Proposed Production-Model of Rapid Subauroral Ion Drifts and Their Relationship to Substorm Evolution, *J. Geophys. Res.*, *98*(A4), 6069-6078.

Angelopoulos, V., et al. (2019), The Space Physics Environment Data Analysis System (SPEDAS), *215*(1), 9.

Artemyev, A. V., R. Rankin, and M. Blanco (2015), Electron trapping and acceleration by kinetic Alfven waves in the inner magnetosphere, *Journal of Geophysical Research: Space Physics*, *120*(12), 10,305-310,316.

Barbier, D. (1958), L'activité aurorale aux basses latitudes, *Ann Geophys*, *14*, 334.

Carlson, H. C., K. Oksavik, and J. I. Moen (2013), Thermally excited 630.0 nm O(1D) emission in the cusp: A frequent high-altitude transient signature, *Journal of Geophysical Research: Space Physics*, *118*(9), 5842-5852.

Chaston, C. C., J. W. Bonnell, C. W. Carlson, J. P. McFadden, R. E. Ergun, and R. J. Strangeway (2003), Properties of small-scale Alfvén waves and accelerated electrons from FAST, *Journal of Geophysical Research: Space Physics*, *108*(A4).

Chaston, C. C., et al. (2014), Observations of kinetic scale field line resonances, *Geophys. Res. Lett.*, *41*(2), 209-215.

Davis, T. N. (1978), Observed Characteristics of Auroral Forms, *Space Sci. Rev.*, *22*(1), 77-113.

Ergun, R. E., Y. J. Su, L. Andersson, C. W. Carlson, J. P. McFadden, F. S. Mozer, D. L. Newman, M. V. Goldman, and R. J. Strangeway (2001), Direct observation of localized parallel electric fields in a space plasma, *Physical Review Letters*, *87*(4).

Evans, J. V., and G. P. Mantas (1968), Thermal structure of the temperate latitude ionosphere, *Journal of Atmospheric and Terrestrial Physics*, *30*(4), 563-577.

Feldstein, Y. I. (1963), Some Problems Concerning the Morphology of Auroras and Magnetic Disturbances at High Latitudes, *Geomagnetism and Aeronomy*, *3*, 183.

Fok, M.-C., J. U. Kozyra, A. F. Nagy, C. E. Rasmussen, and G. V. Khazanov (1993), Decay of equatorial ring current ions and associated aeronomical consequences, *Journal of Geophysical Research: Space Physics*, *98*(A11), 19381-19393.

Förster, M., J. C. Foster, J. Smilauer, K. Kudela, and A. V. Mikhailov (1999), Simultaneous measurements from the Millstone Hill radar and the Active satellite during the SAID/SAR arc event of the March 1990 CEDAR storm, *Ann Geophys-Germany*, *17*(3), 389-404.

Foster, J. C., and W. J. Burke (2002), SAPS: A new categorization for sub-auroral electric fields, *Eos, Transactions American Geophysical Union*, *83*(36), 393-394.

Foster, J. C., and H. B. Vo (2002), Average characteristics and activity dependence of the subauroral polarization stream, *Journal of Geophysical Research: Space Physics*, *107*(A12), SIA 16-11-SIA 16-10.

Foster, J. C., M. J. Buonsanto, M. Mendillo, D. Nottingham, F. J. Rich, and W. Denig (1994), Coordinated stable auroral red arc observations: Relationship to plasma convection, *Journal of Geophysical Research: Space Physics*, *99*(A6), 11429-11439.

Friis-Christensen, E., H. Lühr, G. J. E. Hulot, Planets, and Space (2006), Swarm: A constellation to study the Earth's magnetic field, *58*(4), 351-358.

Funsten, H. O., et al. (2013), Helium, Oxygen, Proton, and Electron (HOPE) Mass Spectrometer for the Radiation Belt Storm Probes Mission, *Space Sci. Rev.*, *179*(1), 423-484.

Gallardo-Lacourt, B., J. Liang, Y. Nishimura, and E. Donovan (2018), On the Origin of STEVE: Particle Precipitation or Ionospheric Skyglow?, *Geophys. Res. Lett.*, *45*(16), 7968-7973.

Gallardo-Lacourt, B., Y. Nishimura, E. Donovan, D. M. Gillies, G. W. Perry, W. E. Archer, O. A. Nava, and E. L. Spanswick (2018), A Statistical Analysis of STEVE, *Journal of Geophysical Research: Space Physics*, *123*(11), 9893-9905.

Galperin, Y. I. (2002), Polarization Jet: Characteristics and a model, *Ann Geophys-Germany*, *20*(3), 391-404.

Gustavsson, B., et al. (2001), First tomographic estimate of volume distribution of HF-pump enhanced airglow emission, *106*(A12), 29105-29123.

He, F., X.-X. Zhang, and B. Chen (2014), Solar cycle, seasonal, and diurnal variations of subauroral ion drifts: Statistical results, *Journal of Geophysical Research: Space Physics*, *119*(6), 5076-5086.

Hoch, R. J., and J. Lemaire (1975), Stable auroral red arcs and their importance for the physics of the plasmapause region, *Ann Geophys*, *31*, 105-110.





Hull, A. J., J. W. Bonnell, F. S. Mozer, and J. D. Scudder (2003), A statistical study of large-amplitude parallel electric fields in the upward current region of the auroral acceleration region, *J. Geophys. Res.*, *108*(A1).

Keiling, A., J. R. Wygant, C. A. Cattell, F. S. Mozer, and C. T. Russell (2003), The Global Morphology of Wave Poynting Flux: Powering the Aurora, *Science*, *299*(5605), 383-386.

Kelley, M. C. (1986), Intense sheared flow as the origin of large-scale undulations of the edge of the diffuse aurora, *Journal of Geophysical Research: Space Physics*, *91*(A3), 3225-3230.

Kivelson, M. G., and C. T. Russell (1995), *Introduction to Space Physics*, 586 pp.

Kletzing, C. A. (1994), Electron acceleration by kinetic Alfvén waves, *Journal of Geophysical Research: Space Physics*, *99*(A6), 11095-11103.

Knudsen, D. J., J. K. Burchill, S. C. Buchert, A. I. Eriksson, R. Gill, J.-E. Wahlund, L. Åhlen, M. Smith, and B. Moffat (2017), Thermal ion imagers and Langmuir probes in the Swarm electric field instruments, *122*(2), 2655-2673.

Kozyra, J. U., C. E. Valladares, H. C. Carlson, M. J. Buonsanto, and D. W. Slater (1990), A theoretical study of the seasonal and solar cycle variations of stable aurora red arcs, *95*(A8), 12219-12234.

Kozyra, J. U., E. G. Shelley, R. H. Comfort, L. H. Brace, T. E. Cravens, and A. F. Nagy (1987), The Role of Ring Current O+ in the Formation of Stable Auroral Red Arcs, *J. Geophys. Res.*, *92*(A7), 7487-7502.

Kurth, W. S., S. De Pascuale, J. B. Faden, C. A. Kletzing, G. B. Hospodarsky, S. Thaller, and J. R. Wygant (2015), Electron densities inferred from plasma wave spectra obtained by the Waves instrument on Van Allen Probes, *J. Geophys. Res.*, *120*(2), 904-914.

Lakhina, G. S., M. Mond, and E. Hameiri (1990), Ballooning mode instability at the plasmapause, *Journal of Geophysical Research: Space Physics*, *95*(A4), 4007-4016.

Liang, J., B. Yang, E. Donovan, J. Burchill, and D. Knudsen (2017), Ionospheric electron heating associated with pulsating auroras: A Swarm survey and model simulation, *Journal of Geophysical Research: Space Physics*, *122*(8), 8781-8807.

Lui, A. T. Y., C.-I. Meng, and S. Ismail (1982), Large amplitude undulations on the equatorward boundary of the diffuse aurora, *Journal of Geophysical Research: Space Physics*, *87*(A4), 2385-2400.

Lysak, R. L., and Y. Song (2003), Kinetic theory of the Alfvén wave acceleration of auroral electrons, *Journal of Geophysical Research: Space Physics*, *108*(A4).

MacDonald, E. A., N. A. Case, J. H. Clayton, M. K. Hall, M. Heavner, N. Lalone, K. G. Patel, and A. Tapia (2015), Aurorasaurus: A citizen science platform for viewing and reporting the aurora, *Space Weather*, n/a-n/a.

MacDonald, E. A., et al. (2018), New science in plain sight: Citizen scientists lead to the discovery of optical structure in the upper atmosphere, *Science Advances*, *4*(3).

Mauk, B., N. Fox, S. Kanekal, R. Kessel, D. Sibeck, and A. Ukhorskiy (2012), Science Objectives and Rationale for the Radiation Belt Storm Probes Mission, *Space Sci. Rev.*, 1-25.

Mishin, E. V. (2013), Interaction of substorm injections with the subauroral geospace: 1. Multispacecraft observations of SAID, *Journal of Geophysical Research: Space Physics*, *118*(9), 5782-5796.

Moffett, R. J., A. E. Ennis, G. J. Bailey, R. A. Heelis, and L. H. J. A. G. Brace (1998), Electron temperatures during rapid subauroral ion drift events, *16*(4), 450-459.

Mozer, F. S. (1970), Electric field mapping in the ionosphere at the equatorial plane, *Planetary and Space Science*, *18*(2), 259-263.

Mozer, F. S., C. W. Carlson, M. K. Hudson, R. B. Torbert, B. Parady, J. Yatteau, and M. C. Kelley (1977), Observations of Paired Electrostatic Shocks in Polar Magnetosphere, *Physical Review Letters*, *38*(6), 292-295.

Ni, B., R. M. Thorne, Y. Y. Shprits, and J. Bortnik (2008), Resonant scattering of plasma sheet electrons by whistler-mode chorus: Contribution to diffuse auroral precipitation, *Geophys. Res. Lett.*, *35*(11), L11106.

Nishimura, Y., B. Gallardo-Lacourt, Y. Zou, E. Mishin, D. J. Knudsen, E. F. Donovan, V. Angelopoulos, and R. Raybell (2019), Magnetospheric Signatures of STEVE: Implications for the Magnetospheric Energy Source and Interhemispheric Conjugacy, *Geophys. Res. Lett.*, *0*(0).

Puhl-Quinn, P. A., H. Matsui, E. Mishin, C. Mouikis, L. Kistler, Y. Khotyaintsev, P. M. E. Décréau, and E. Lucek (2007), Cluster and DMSP observations of SAID electric fields, *Journal of Geophysical Research: Space Physics*, *112*(A5).

Rees, M. (1964), Note on the penetration of energetic electrons into the earth's atmosphere, *Planetary and Space Science*, *12*(7), 722-725.

Rees, M., and R. G. Roble (1975), Observations and theory of the formation of stable auroral red arcs, *Rev Geophys*, *13*(1), 201-242.

Ritter, P., H. Lühr, J. J. E. Rauberg, Planets, and Space (2013), Determining field-aligned currents with the Swarm constellation mission, *65*(11), 9.





Sazykin, S., B. G. Fejer, Y. I. Galperin, L. V. Zinin, S. A. Grigoriev, and M. Mendillo (2002), Polarization jet events and excitation of weak SAR arcs, *Geophys. Res. Lett.*, *29*(12).

Shiokawa, K., Y. Miyoshi, P. C. Brandt, D. S. Evans, H. U. Frey, J. Goldstein, and K. Yumoto (2013), Ground and satellite observations of low-latitude red auroras at the initial phase of magnetic storms, *J. Geophys. Res.*, *118*(1), 256-270.

Slater, D. W., C. Gurgiolo, J. U. Kozyra, E. W. Kleckner, and J. D. Winningham (1987), A possible energy source to power stable auroral red arcs: Precipitating electrons, *Journal of Geophysical Research: Space Physics*, *92*(A5), 4543-4552.

Solomon, S. C. (2017), Global modeling of thermospheric airglow in the far ultraviolet, *J. Geophys. Res.*, *122*(7), 7834-7848.

Southwood, D. J., and R. A. Wolf (1978), An assessment of the role of precipitation in magnetospheric convection, *Journal of Geophysical Research: Space Physics*, *83*(A11), 5227-5232.

Spiro, R. W., R. A. Heelis, and W. B. Hanson (1977), Implications of Penetration of Magnetospheric Convection Electric-Field into Mid-Latitude Trough, *Eos T Am Geophys Un*, *58*(8), 754-754.

Thorne, R. M., B. Ni, X. Tao, R. B. Horne, and N. P. Meredith (2010), Scattering by chorus waves as the dominant cause of diffuse auroral precipitation, *Nature*, *467*(7318), 943-946.

Tian, S. (2017), The Flow of Poynting Flux into the Terrestrial Cusp and Auroral Zone and Its Role In Powering Energy Intensive Collisionless Acceleration Mechhanisms.

Unick, C. W., E. Donovan, M. Connors, and B. Jackel (2017), A dedicated H-beta meridian scanning photometer for proton aurora measurement, *122*(1), 753-764.

Vasyliunas, V. (1970), Mathematical Models of Magnetospheric Convection and its Coupling to the Ionosphere, in *Particles and Fields in the Magnetosphere*, edited by B. M. McCormac, pp. 60-71, Springer Netherlands.

Wolf, R. A. (1983), The Quasi-Static (Slow-Flow) Region of the Magnetosphere, Springer Netherlands, Dordrecht.

Wolf, R. A. (1995), Magnetospheric configuration, in *Introduction to Space Physics*, edited by M. G. Kivelson and C. T. Russell.

Wygant, J. R., et al. (2002), Evidence for kinetic Alfven waves and parallel electron energization at 4-6 R-E altitudes in the plasma sheet boundary layer, *J. Geophys. Res.*, *107*(A8).

Yahnin, A. G., V. A. Sergeev, B. B. Gvozdevsky, and S. Vennerstrøm (1997), Magnetospheric source region of discrete auroras inferred from their relationship with isotropy boundaries of energetic particles, *Ann. Geophys.*, *15*(8), 943-958.

Yeh, H. C., J. C. Foster, F. J. Rich, and W. Swider (1991), Storm time electric field penetration observed at mid-latitude, *Journal of Geophysical Research: Space Physics*, *96*(A4), 5707-5721.




# Identifying the magnetospheric driver of STEVE


**Xiangning Chu[1], David Malaspina[1], Bea Gallardo-Lacourt[2], Jun Liang[2], Laila Andersson[1], Qianli Ma[3], Anton Artemyev[4], Jiang Liu[3], Bob Ergun[1], Scott Thaller[1], Hassanali Akbari[1], Hong Zhao[1], Brian Larsen[5], Geoffrey Reeves[5], John Wygant[6], Aaron Breneman[6], Sheng Tian[6], Martin Connors[7], Eric Donovan[2], William Archer[8], Elizabeth A. MacDonald[9]**

[1] Laboratory for Atmospheric and Space Physics, University of Colorado Boulder, Boulder, Colorado, USA

[2] Department of Physics and Astronomy, University of Calgary, Calgary, Alberta, Canada

[3] Department of Atmospheric and Oceanic Sciences, University of California, Los Angeles, California, USA

[4] Institute of Geophysics and Planetary Physics, University of California, Los Angeles, CA, USA

[5] Los Alamos National Laboratory, Los Alamos, NM, United States

[6] Department of Physics, University of Minnesota, Twin Cities, Minneapolis, MN, USA

[7] Athabasca University, Athabasca, Canada

[8] Department of Physics and Engineering Physics, University of Saskatchewan, Saskatoon, Saskatchewan, Canada

[9] NASA Goddard Space Flight Center, Greenbelt, MD, USA




Corresponding author: Xiangning Chu ([chuxiangning@gmail.com](mailto:chuxiangning@gmail.com))

**Contents of this file**

> Move S1, S2 and Figures S1, S2, S3

**Introduction**

Movie S1, S2, and Figure S1, S2, S3 complementing information reported in the main text.

**Movie S1.** STEVE event observed on 17 July 2018 by the THEMIS ASI at ATHA and PINA, with the footprints of Van Allen Probes (blue for A and yellow for B) and Swarm-B (red).

**Movie S2.** STEVE event observed on 17 July 2018 by the Lucky Lack REGO ASI, with the footprints of Van Allen Probes (blue for A and yellow for B) and Swarm-B (red).

**Figure S1.** The STEVE event observed on 17 July 2018 by the REGO ASI at LUCK at (a) 06:30:00 UT, (b) 06:36:00 UT, (c) 06:47:00 UT and (d) 06:53:00 UT, with the footprints of Van Allen Probes (blue for A and yellow for B) and Swarm-B (red).

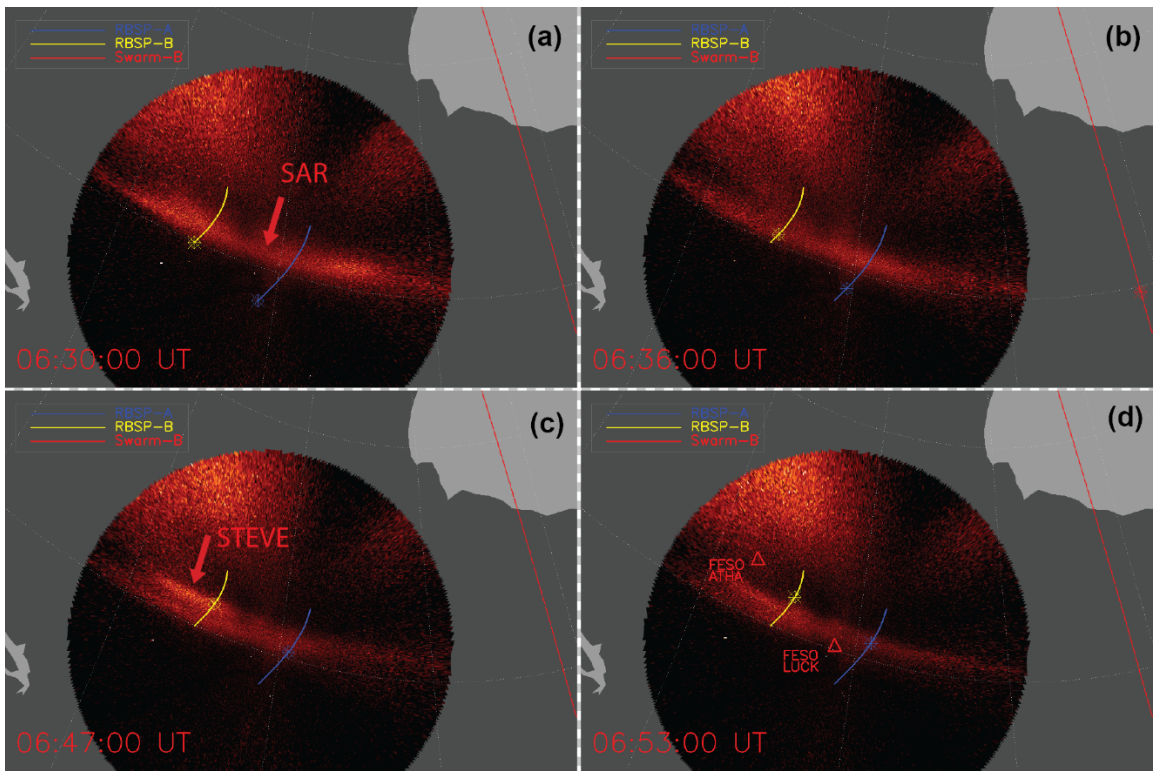



**Figure S2.** The STEVE arc on 17 July 2018 was captured by two sky watchers (Colin Chatfield and Neil Zeller).

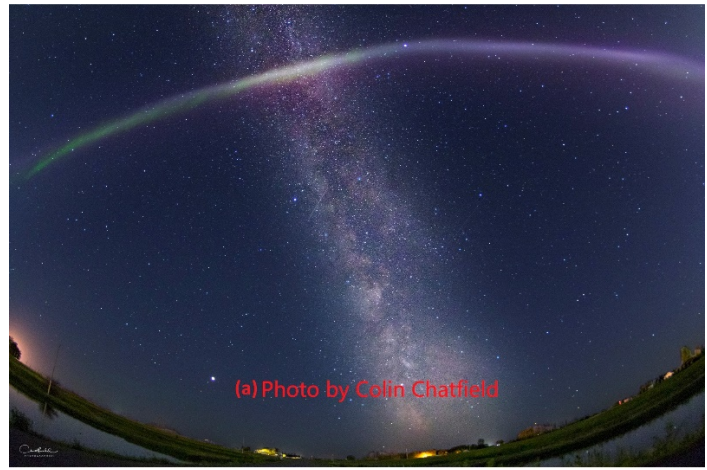

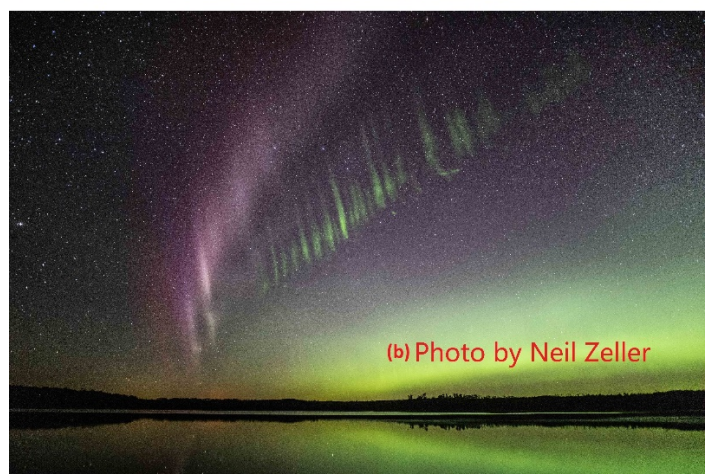



**Figure S3.** Keograms of the proton aurora observed by the Forty-Eight Sixty-One (FESO) meridian scanning photometers (MSPs) at LUCK (geographic coordinates: 252.74 °, 51.15º) and ATHA (geographic coordinates: 246.69º, 54.71º) that record proton aurora emission at 486.1 nm. Their locations are marked in Figure S1 (d). The MSP at LUCK was located close to STEVE, while the proton aurora was located 4º poleward than the center of the MSP at luck (51.15º). The MSP at ATHA was located poleward of STEVE, and the proton aurora was further poleward than its center (54.71º). Therefore, the proton aurora was a few degrees poleward of STEVE throughout the event.

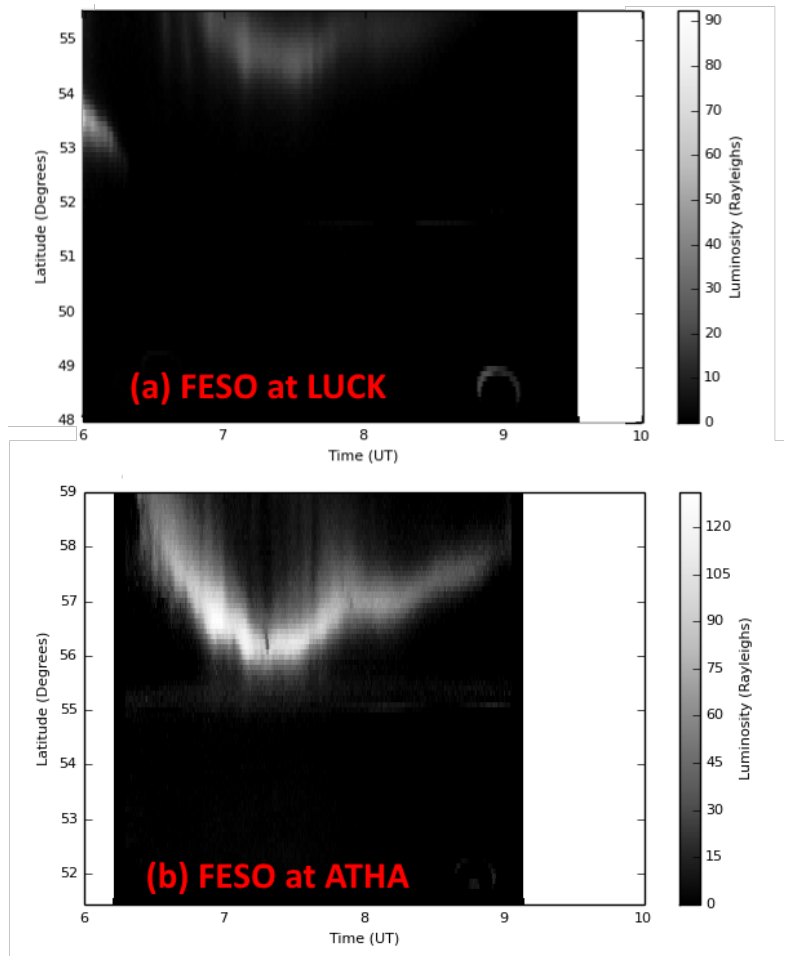